\documentclass[usenatbib,lscape]{emulateapj}
\usepackage{graphicx,ifthen,url,float,lscape,color,ulem}
\newcommand {\kms}{km s$^{-1}$}

\newcommand{\scubaii}{{\sc Scuba-2}}
\newcommand{\clname}{PCL1002}

\newcommand {\lya}{Ly$\alpha$}
\def\ltsima{$\; \buildrel < \over \sim \;$}
\def\simlt{\lower.5ex\hbox{\ltsima}}
\def\gtsima{$\; \buildrel > \over \sim \;$}
\def\simgt{\lower.5ex\hbox{\gtsima}}
\newcommand {\uJy}{$\mu$Jy}
\newcommand {\um}{$\mu$m}

\def\um     {$\mu$m}
\def\ts     {\thinspace}
\def\kms    {\ifmmode{{\rm \ts km\ts s}^{-1}}\else{\ts km\ts s$^{-1}$}\fi}
\def\msol   {\ifmmode{{\rm M}_{\odot}}\else{M$_{\odot}$}\fi}
\def\lsol   {\ifmmode{{\rm L}_{\odot}}\else{L$_{\odot}$}\fi}
\def\zsol   {\ifmmode{{\rm Z}_{\odot}}\else{Z$_{\odot}$}\fi}
\def\etal   {{\rm et\ts al.}}
\def\ci     {\ifmmode{{\rm C}{\rm \small I}}\else{C\ts {\scriptsize I}}\fi}
\def\hi     {\ifmmode{{\rm H}{\rm \small I}}\else{H\ts {\scriptsize I}}\fi}
\def\hh     {\ifmmode{{\rm H}_2}\else{H$_2$}\fi}
\def\cone {\ifmmode{{\rm C}{\rm \small I}(^3\!P_1\!\to^3\!P_0)}
     \else{C\ts {\scriptsize I}{\small$(^3\!P_1\!\to\,^3\!P_0)$}}\fi}
\def\ctwo {\ifmmode{{\rm C}{\rm \small I}(^3\!P_2\!\to\,^3\!P_1)}
     \else{C\ts {\scriptsize I}{\small$(^3\!P_2\!\to\,^3\!P_1)$}}\fi}
\def\cij {\ifmmode{{\rm C}{\rm \small I}\,(^3P_i\to^3P_j)}\else{C\ts {\scriptsize I}\,{\small$(^3P_i\to^3P_j)$}}\fi}
\def\cii    {\ifmmode{{\rm C}{\rm \small II}}\else{C\ts {\scriptsize II}}\fi}
\def\tex {\ifmmode{{T}_{\rm ex}}\else{$T_{\rm ex}$}\fi}
\def\tmb {\ifmmode{{T}_{\rm mb}}\else{$T_{\rm mb}$}\fi}
\def\tkin {\ifmmode{{T}_{\rm kin}}\else{$T_{\rm kin}$}\fi}
\def\microns {\ifmmode{\mu{\rm m}}\else{$\mu$m}\fi}
\def\nhh   {\ifmmode{n({\rm H}_2)}\else{$n$(H$_2$)}\fi}

\newcommand{\msun}{{\rm\,M$_\odot$}}
\newcommand{\sfr}{{\rm\,M$_\odot$\,yr$^{-1}$}}
\newcommand{\lsun}{{\rm\,L$_\odot$}}

\newcommand{\ha}{{\rm\,H$\alpha$}}

\shorttitle{A DSFG-rich, AGN-rich $z=2.47$ proto-cluster}
\shortauthors{C.~M. Casey et al.}
\begin{document}

\title{A Massive, Distant Proto-cluster at $z=2.47$ Caught in a Phase of Rapid Formation?}

\author{C.M.~Casey\altaffilmark{1,2}, A. Cooray\altaffilmark{2}, P. Capak\altaffilmark{3},
H. Fu\altaffilmark{4}, K. Kovac\altaffilmark{5}, S. Lilly\altaffilmark{5}, D.B. Sanders\altaffilmark{6}, 
N.Z. Scoville\altaffilmark{7}, E. Treister\altaffilmark{8}}
\altaffiltext{1}{Department of Astronomy, the University of Texas at Austin, 2515 Speedway Blvd, Stop C1400, Austin, TX 78712}
\altaffiltext{2}{Department of Physics and Astronomy, University of California, Irvine, Irvine, CA 92697}
\altaffiltext{3}{Infrared Processing and Analysis Center (IPAC), 1200 E. California Blvd., Pasadena, CA 91125}
\altaffiltext{4}{Department of Physics and Astronomy, University of Iowa, 751 Van Allen Hall, Iowa City, IA 52242}
\altaffiltext{5}{Institute for Astronomy, ETH Zurich, CH-8093 Zurich, Switzerland}
\altaffiltext{6}{Institute for Astronomy, University of Hawai'i, 2680 Woodlawn Dr, Honolulu, HI 96822}
\altaffiltext{7}{California Institute of Technology, 1216 East California Boulevard, Pasadena, CA 91125}
\altaffiltext{8}{Departamento de Astronom\'{i}a, Universidad de Concepci\'{o}n, Av. Esteban Iturra, Concepci\'{o}n, Chile}

\label{firstpage}

\begin{abstract}
Numerical simulations of cosmological structure formation show that
the Universe's most massive clusters, and the galaxies living in those
clusters, assemble rapidly at early times ($2.5<z<4$).  While more
than twenty proto-clusters have been observed at $z\simgt2$ based on
associations of 5--40 galaxies around rare sources, the observational
evidence for rapid cluster formation is weak.  Here we report
observations of an asymmetric, filamentary structure at $z=2.47$
containing seven starbursting, submillimeter-luminous galaxies and
five additional AGN within a comoving volume of 15000\,Mpc$^3$.  As
the expected lifetime of both the luminous AGN and starburst phase of
a galaxy is $\sim$100\,Myr, we conclude that these sources were likely
triggered in rapid succession by environmental factors or,
alternatively, the duration of these cosmologically rare phenomena is
much longer than prior direct measurements suggest.  The stellar mass
already built up in the structure is $\sim$10$^{12}$\,\msun\ and we
estimate that the cluster mass will exceed that of the Coma
supercluster at $z\sim0$.  The filamentary structure is in line with
hierarchical growth simulations which predict that the peak of cluster
activity occurs rapidly at $z>2$.
\end{abstract}

\keywords{
galaxies: clusters: general $-$ galaxies: starburst $-$ galaxies: quasars: general $-$ cosmology: large-scale structure of universe
}

\section{Introduction}

An outstanding question on the study of massive galaxy clusters in the
Universe is how and when the member galaxies built most of their mass:
in a cascade before the cluster coalesces, gradually as the cluster
accretes mass, or predominantly after the formation of the cluster.
While galaxies in nearby coalesced clusters have suppressed
star-formation rates due to ``cluster quenching,'' \citep{cooper08a}
it is unclear whether or not a reversal occurs at
high-redshift \citep{elbaz07a}, whereby galaxies in more massive
structures have enhanced star-formation rates, in line with
expectation from hierarchical growth formation \citep{moster13a}.
Dense environments that are undergoing a rapid formation in the form
of proto-clusters are difficult to detect because the intracluster
medium at $z>2.5$ has not yet been heated sufficiently to emit in the
X-rays or absorb cosmic microwave background photons via the
Sunyaev-Zel'dovich effect.  The existing discoveries of high-redshift
dense structures do not provide adequate observational evidence to
interpret how and when the galaxies in those regions formed.

Here we present data on a distant proto-cluster at $z=2.47$ found
serendipitously during a redshift survey of dusty star-forming
galaxies (DSFGs) in the \scubaii-imaged portion of the COSMOS field
\citep{casey13a}.  We discuss data and observations in
\S~\ref{sec:data}, present relevant results and calculations in
\S~\ref{sec:results}, and discuss the implications on the formation of
early proto-clusters in \S~\ref{sec:conclusions}.

\section{Data \&\ Observations}\label{sec:data}

\begin{figure*}
\centering
\vspace{-8mm}
\includegraphics[width=0.9\columnwidth]{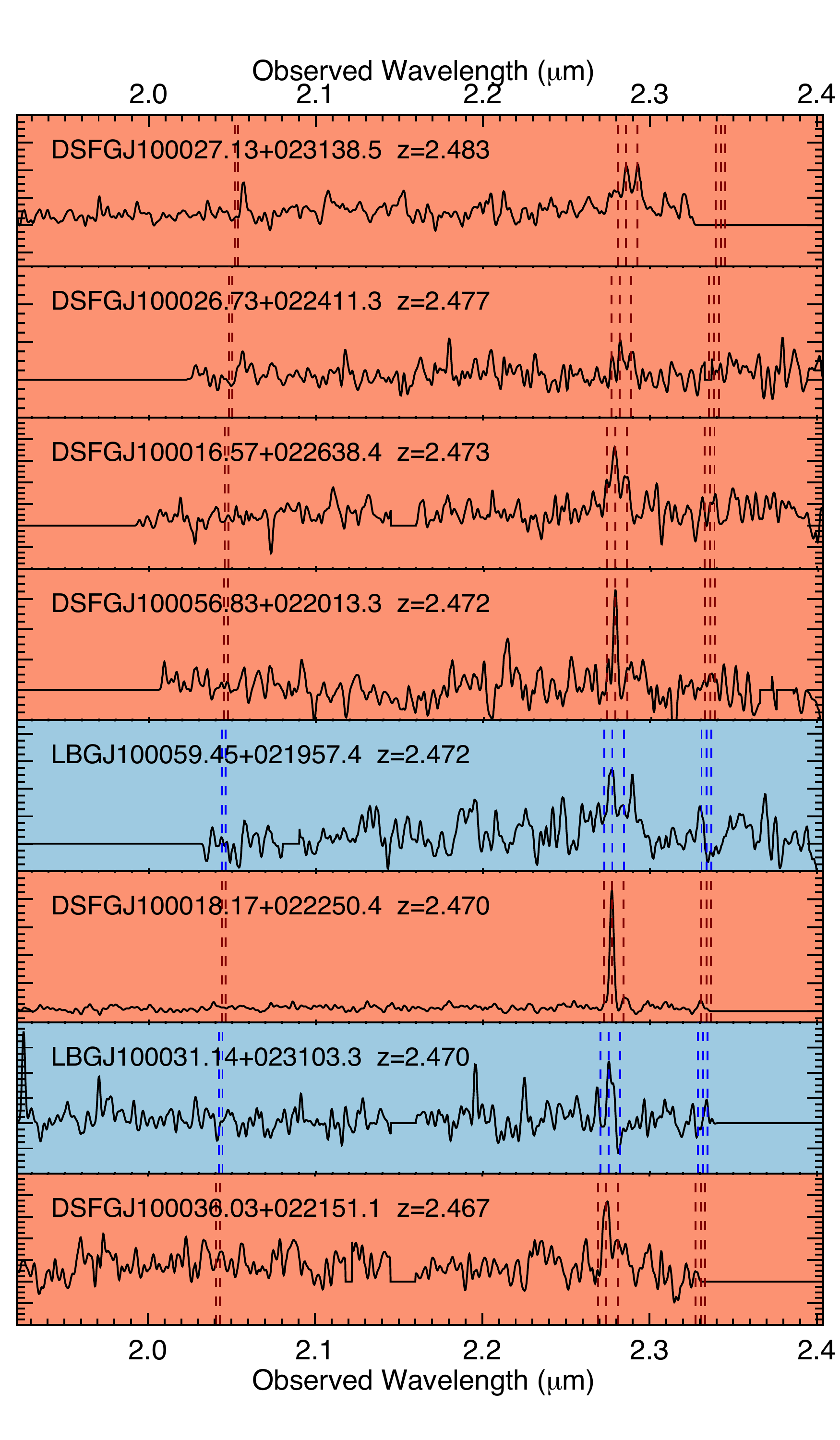}
\includegraphics[width=1.0\columnwidth]{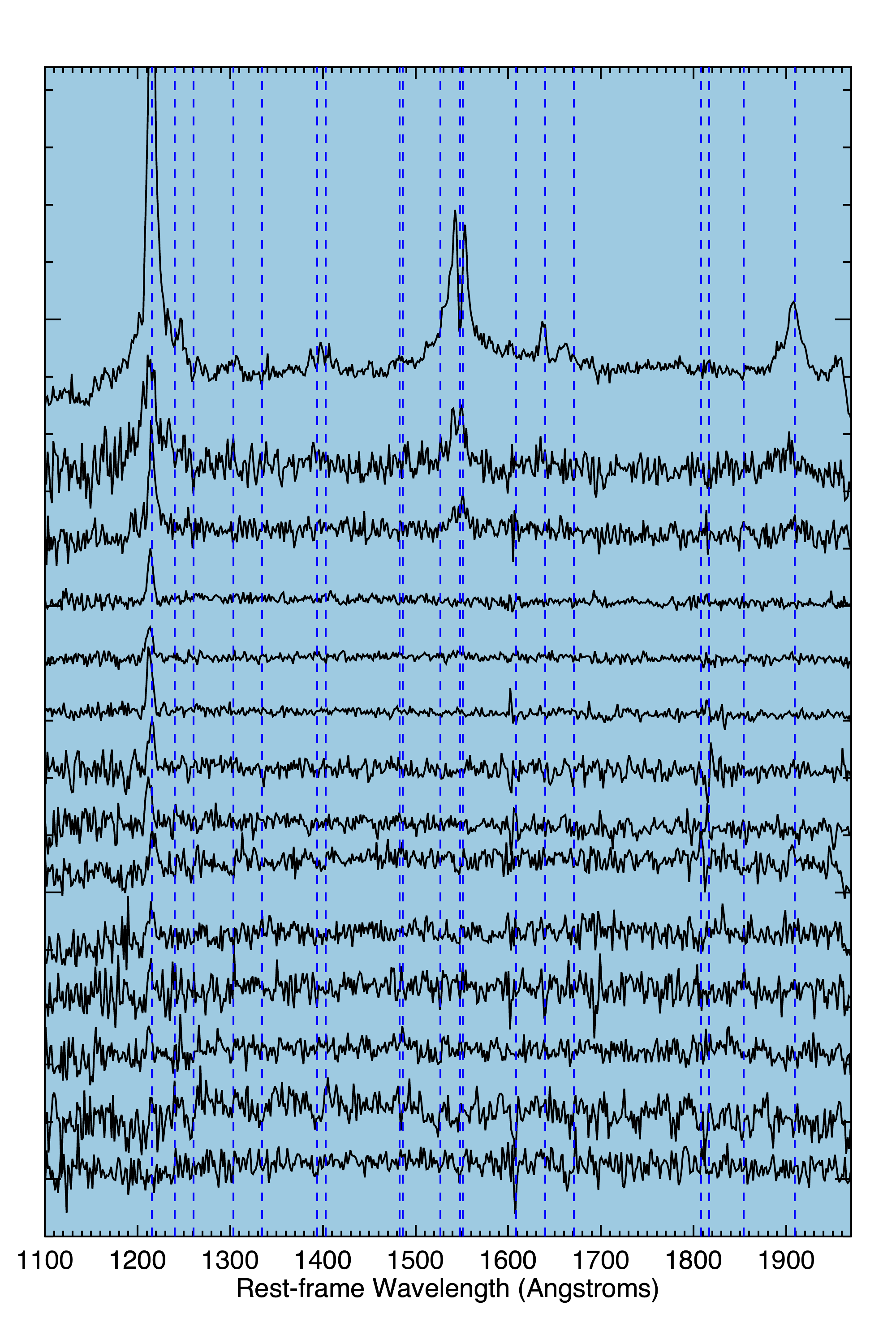}
\caption{Left: extracted one-dimensional K-band MOSFIRE spectra
  for proto-cluster members. 
  Right: example spectra of \clname\ member galaxies in the rest-frame
  ultraviolet from VLT VIMOS \citep{lilly09a}. 
}
\label{fig:spectra}
\end{figure*}

This structure, which we call \clname, was found serendipitously in Keck
MOSFIRE (21-Dec-2012, 31-Dec-2013 and 19-Jan-2014) spectroscopic
follow-up of \scubaii-selected DSFGs in the COSMOS field. The DSFGs'
FIR-photometry is given in Table~\ref{tab:phot}. Observing conditions
for MOSFIRE nights were favorable, with clear skies and
0.5--0.7\arcsec\ seeing.  Six DSFGs were spectroscopically confirmed
with \ha\ redshifts within $\pm$0.007 of $z=2.472$.  The DSFGs'
near-infrared counterparts are very secure for 4/6 DSFGs, driven by
precise 450\um\ positions \citep[all given in][except
  DSFG\,J100026.73+022411.3 with $S_{\rm
    450}=14.6\pm4.1$\,mJy]{casey13a}.  Two others have more ambiguity
due to 850\um-selection.  DSFGJ100018.17$+$022250.4 resembles a
major-merger spanning 2$''$ with multiple
knots. DSFGJ100027.14$+$023140.8 has both 24\um/radio emission
overlapping with IRAC emission towards the identified counterpart,
given by source name.  Two other MOSFIRE targets were confirmed in
this same redshift interval.  The one-dimensional \ha\ spectra are
shown in Figure~\ref{fig:spectra}.

\begin{table*}
\caption{Deboosted FIR-Photometric data for \clname's dusty starbursts}
\vspace{-4mm}
\begin{center}
{\footnotesize
\begin{tabular}{l@{ }c@{ }c@{ }c@{ }c@{ }c@{ }c@{ }c@{ }c@{ }c@{ }c@{ }c@{ }c}
\hline\hline
{\sc Name} & $S_{24}$ & $S_{100}$ & $S_{160}$ & $S_{250}$ & $S_{350}$ & $S_{450}$ & $S_{500}$ & $S_{850}$  & $S_{\rm 1.4GHz}$ & $L_{\rm IR}$\\
           & [\uJy]     & [mJy]     & [mJy]     & [mJy]     & [mJy]     & [mJy]     & [mJy]     & [mJy]     & [\uJy]           & [\lsun] \\
\hline
%DSFG450.58
DSFG\,J100036.03+022151.1... & 194$\pm$17 & $-$          & $-$           & 11.3$\pm$2.2 & 15.6$\pm$2.7 & 11.3$\pm$4.9 & 14.5$\pm$3.1 & 4.6$\pm$1.1  & $-$  & (3.12$^{+1.39}_{-0.96}$)$\times$10$^{12}$ \\
%DSFG850.44
DSFG\,J100018.17+022250.4... & 128$\pm$16 & $-$          & $-$           & 11.3$\pm$2.2 & $-$          & 3.9$\pm$4.1  & $-$          & 3.3$\pm$1.0  & $-$  & (3.84$^{+2.35}_{-1.46}$)$\times10^{12}$ \\
%DSFG450.09
DSFG\,J100016.57+022638.4... & 890$\pm$17 & 6.7$\pm$1.9  & 19.2$\pm$3.6  & 24.5$\pm$2.2 & 21.5$\pm$2.7 & 17.3$\pm$4.7 & 11.6$\pm$3.0 & 3.7$\pm$1.0  & 5716$\pm$73 & (7.52$^{+2.21}_{-1.71}$)$\times$10$^{12}$ \\
%DSFG450.28
DSFG\,J100056.83+022013.3... & 90$\pm$27  & $-$          & $-$           & $-$          & 9.1$\pm$12.9 & 18.3$\pm$6.0 & $-$          & 10.9$\pm$1.1 & $-$   & (2.06$^{+0.93}_{-0.64}$)$\times10^{12}$ \\
%SPIRE457
DSFG\,J100026.73+022411.3... &  84$\pm$14 & $-$ & $-$ & 11.4$\pm$2.2 & 14.8$\pm$2.7 & 10.3$\pm$5.0 & 17.8$\pm$3.0 & 0.4$\pm$0.8 & $-$ & (2.99$^{+1.85}_{-1.16}$)$\times10^{12}$ \\
%COLDz
DSFG\,J100018.21+023456.7... & 153$\pm$11 & $-$          & $-$           & 16.6$\pm$2.0 & 19.5$\pm$3.4 & 12.0$\pm$8.8  & 11.2$\pm$4.2 & 2.57$\pm$1.74 & 46$\pm$10 & (4.23$^{+4.51}_{-2.18}$)$\times10^{12}$ \\
%850.20
DSFG\,J100027.14+023140.8... & 421$\pm$152 & 13.6$\pm$1.8 & 24.4$\pm$3.6 & 36.9$\pm$2.2 & 30.8$\pm$2.8 & 3.9$\pm$5.0  & 17.0$\pm$3.4 & 5.8$\pm$1.4 & 67$\pm$12 & (1.15$^{+0.19}_{-0.16}$)$\times10^{13}$ \\
\hline\hline
\end{tabular}
}
\end{center}
\label{tab:phot}
\end{table*}

One additional {\it Herschel}-{\sc SPIRE}-detected galaxy,
COLDz\,J100018.21$+$023456.7, sits at $z=2.4790$ \citep{lentati15a},
confirmed via detection of CO(1-0) in a 6.5\,arcmin$^2$ blank-field CO
search program north of \clname\ (D.~Riechers \etal, in~prep).

Supplementary data are pooled from a repository of legacy ancillary
data in COSMOS.  An additional 34 spectroscopically-confirmed
sources in the $z$COSMOS survey are within $2.463<z<2.487$ 
\citep{lilly09a}, identified via \lya\ emission, Fe{\sc ii}, Si{\sc
  ii} and C{\sc ii} absorption, also shown in
Figure~\ref{fig:spectra}.  We also make use of the 30$+$ photometric
bands of imaging data available in the field \citep{ilbert13a}.  We
also draw on the COSMOS {\it Chandra} X-ray 0.5--10\,keV catalog
\citep{civano12a}, radio 1.4\,GHz catalog \citep{schinnerer07a}, and
      {\it Herschel} PEP/PACS and HerMES/SPIRE 100--500\um\ catalogs
      \citep{lee13a}.

We calculate the significance of this over-density using methods used
for other $z\simgt2$ structures \citep{steidel98a,chapman09a} by
computing the likelihood of observing 7 DSFGs within a
$\Delta z=0.02$ interval with the Erlang distribution function
\citep{eadie71a},
\begin{equation}
p(\Delta\!z|N\lambda) = \lambda(\lambda\Delta\!z)^{N-2}\exp(-\lambda\Delta\!z)/(N-2)!
\end{equation}
 where $p(\Delta\!z|N\lambda)$ is the probability that $N$ galaxies
 spanning a redshift range $\Delta z$ is drawn by chance.  The
 expectation for the density of galaxies per unit redshift interval is
 described by $\lambda$ and the calculation of the probability assumes
 no clustering as the null hypothesis.  Assuming a volume-density of
 7$\times$10$^{-5}$\,Mpc$^{-3}$ for $>10^{12.3}$\,\lsun\ DSFGs at
 $z\approx2.5$ \citep*[taken from the best to-date luminosity
   functions of DSFGs;][]{casey14a}, we infer that the number of DSFGs
 expected in a redshift slice of $\Delta z=0.02$ within a
 150\,arcmin$^2$ box is $\lambda=0.64$ (in a volume
 $\sim$10000\,Mpc$^3$).  This implies the probability of observing
 seven DSFGs in this interval is 0.002\%.  This corresponds to a DSFG
 over-density of $\delta_{\rm DSFG}=(7-0.64)/0.64=10$. For comparison,
 the most analogous structures in the literature are in
 HDF at $z=1.99$ \citep[][ which has $\delta_{\rm
     DSFG}=(9-0.84)/0.84=10$]{blain04a}, and the SSA22 proto-cluster
 at $z=3.09$ with $\approx$5--6\,DSFGs.  MRC1138$-$256 at $z=2.16$
 \citep{dannerbauer14a} contains $\sim$5 DSFGs, within a potentially
 much more compact volume ($\delta_{\rm DSFG}\simgt100$).
With only 0.6 DSFGs expected in the given volume, this over-density
only has a 0.0001\%\ chance of occurring by chance, implying that
\clname\ is unlikely to be an artifact of incompleteness or survey
bias.

\begin{figure*}
\centering
\includegraphics[width=2.0\columnwidth]{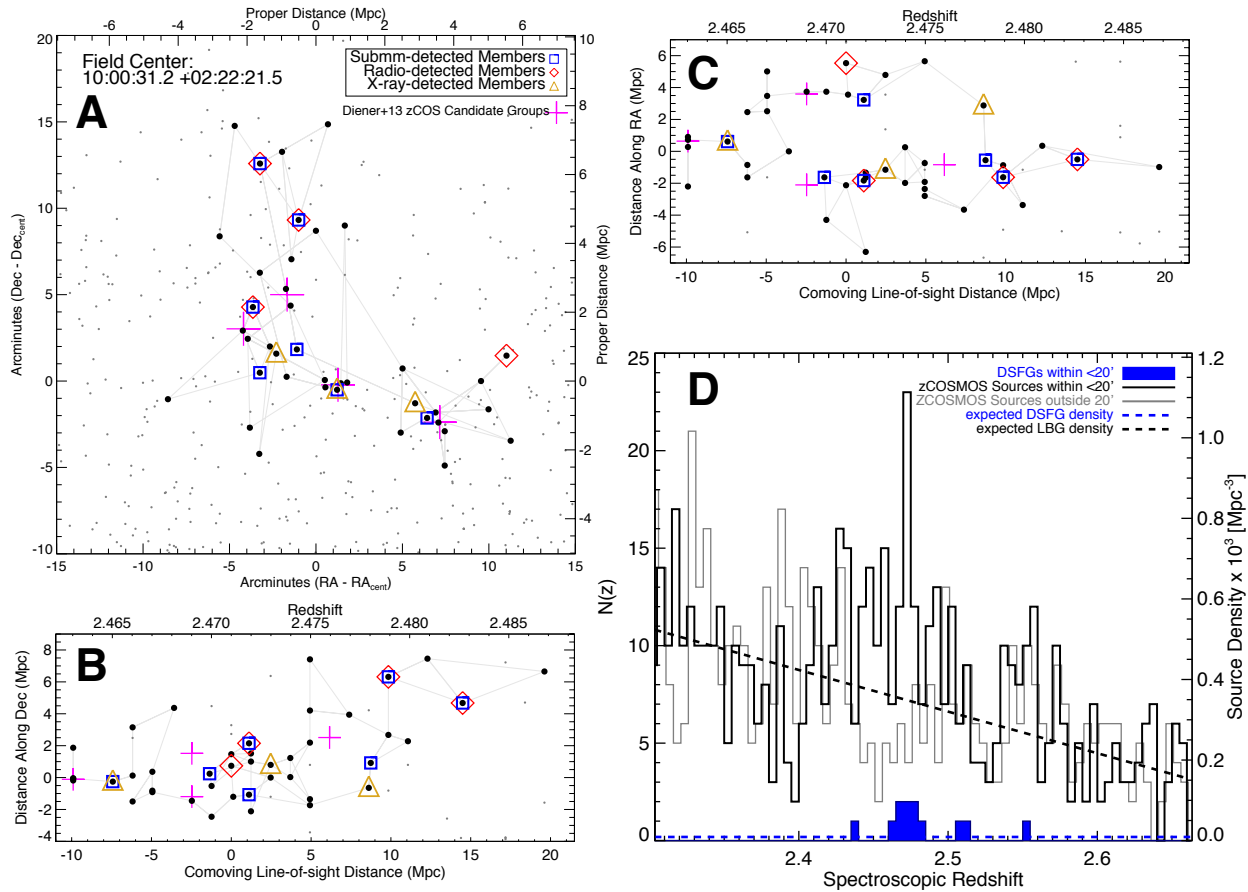}
\vspace{-8mm}
\caption{ Three spatial projections of the $z=2.47$
  \clname\ proto-cluster: on the sky (panel A), and two projections
  with redshift on the x-axis and either decl.  (panel B) or
  R.A. (panel C) on the y-axis.  Thick black points are confirmed
  \clname\ members, and small gray points are
  spectroscopically-confirmed galaxies in $2.30<z<2.66$.  Thin gray
  lines connect the two nearest-neighbors.  DSFGs have blue boxes,
  radio-detected galaxies have red diamonds, and X-ray-detected
  galaxies have yellow triangles.  Peculiar velocities are not
  constrained.  Four candidate group positions are marked with large
  magenta crosses \citep{diener13a}.  Panel D shows the distribution
  of spectroscopic redshifts for galaxies in within 20\arcmin\ of
  \clname's center.  The over-density of dusty starburst,
  submillimeter galaxies (DSFGs; blue) is a factor of 10 above what is
  expected (dashed blue line) given the known volume density of DSFGs
  in the field at these redshifts. The observed number of LBGs (dashed
  black line) is a factor of 3.3 above expectation at $z=2.47$.}
\label{fig:map}
\end{figure*}

A similar peak is seen in the redshift distribution of $z$COSMOS
Lyman-Break Galaxies (LBGs) with a maximum $\delta_{\rm g}=3.3$.
Although less pronounced than the DSFG over-density, the over-density
of Lyman Break Galaxies are also spatially clustered on the sky in a
filamentary structure. We use a friends-of-friends
\citep[FOF;][]{huchra82a} algorithm to formally determine which LBGs
are in fact members of the DSFG over-density by selecting sources
which are within 2\,Mpc proper (projected on the sky and
line-of-sight distance) of the DSFGs or their immediate neighbors, or
within 3\,Mpc of the DSFGs.  This is similar to low-$z$
cluster member identification techniques but with longer `linking
length' accommodating high-$z$ non-virialized structures 
\citep{chiang13a}.
Table~\ref{tab:allmems} lists \clname\ members.

\begin{table*}
\caption{Physical Characteristics of \clname\ Members}
\begin{center}
{\footnotesize
\vspace{-5mm}
\begin{tabular}{l@{ }c@{ }c@{ }c@{ }c@{ }c@{ }c@{ }c@{ }c@{ }c}
\hline\hline
{\sc Name} & z & {\sc Type} & L$_{\rm UV}$ & SFR                  & M$_\star$   & {\sc Morph}  & {\sc AGN} \\
           &   &            & (L$_\odot$)  & (\msun\,yr$^{-1}$) & (M$_\odot$)   &  {\sc Class} &  {\sc indicator}\\
\hline
%%ZCOS01
LBG\,J100013.62+022604.9 & 2.463 & UV & (4.26$^{+0.66}_{-0.57}$)$\times10^{10}$   & 6.67$^{+0.55}_{-1.41}$ & (1.14$^{+0.06}_{-0.21}$)$\times10^{10}$ & Disk/Int & $-$ \\
%ZCOS16
LBG\,J100033.33+022159.9 & 2.463 & UV & (4.36$^{+0.69}_{-0.60}$)$\times10^{10}$   & 49.5$^{+0.6}_{-21.1}$  & (4.65$^{+0.52}_{-0.87}$)$\times10^{10}$ & Sph & $-$ \\
%ZCOS17
LBG\,J100036.90+022213.8 & 2.463 & \lya/UV & (2.24$^{+0.07}_{-0.07}$)$\times10^{10}$   & 3.17$^{+0.20}_{-1.02}$ & (5.74$^{+0.90}_{-0.90}$)$\times10^{8}$ & Sph & $-$ \\
%ZCOS18
LBG\,J100038.35+022216.4 & 2.463 & UV & (5.86$^{+0.78}_{-0.69}$)$\times10^{10}$   & 25.9$^{+29.1}_{-3.5}$  & (7.92$^{+1.01}_{-2.09}$)$\times10^{9}$ & Sph/Int & $-$ \\

%DSFG450.58
DSFG\,J100036.03+022151.1 & 2.465 & \ha\ &       $<$3.0$\times$10$^{10}$             & 296$^{+132}_{-91}$     & (1.01$^{+0.06}_{-0.18}$)$\times10^{11}$ & Disk/Int & X-ray \\
%ZCOS06
LBG\,J100018.18+022837.7 & 2.466 & UV & (4.28$^{+0.69}_{-0.59}$)$\times10^{10}$   & 12.7$^{+3.4}_{-2.7}$   & (2.37$^{+0.30}_{-0.43}$)$\times10^{9}$ & Sph/Int & $-$ \\
%ZCOS13
LBG\,J100024.36+022236.3 & 2.466 & UV & (4.88$^{+0.70}_{-0.61}$)$\times10^{10}$   & 25.5$^{+4.8}_{-9.8}$   & (8.79$^{+0.08}_{-0.09}$)$\times10^{9}$ & Sph/Int & $-$ \\
%ZCOS19
LBG\,J100050.73+021922.4 & 2.466 & UV & (6.61$^{+0.81}_{-0.72}$)$\times10^{10}$   & 98.2$^{+7.6}_{-17.4}$  & (1.44$^{+0.06}_{-0.28}$)$\times10^{10}$ & $-$ & $-$ \\

%COMP422
LBG\,J100031.14+023103.3  & 2.467 & \ha/\lya\ & (2.61$^{+0.66}_{-0.59}$)$\times10^{10}$   & 8.21$^{+1.96}_{-1.96}$ & (3.22$^{+0.20}_{-0.31}$)$\times10^{10}$ & Sph & $-$ \\
%ZCOS20
LBG\,J100051.16+022305.1 & 2.467 & \lya\ & (2.86$^{+0.63}_{-0.51}$)$\times10^{10}$   & 17.8$^{+0.9}_{-7.3}$   & (2.01$^{+0.15}_{-1.03}$)$\times10^{9}$ & $-$ & $-$ \\
%ZCOS22
LBG\,J100058.80+022032.4 & 2.467 & UV & (4.90$^{+0.74}_{-0.64}$)$\times10^{10}$   & 63.1$^{+1.2}_{-8.2}$   & (5.94$^{+0.57}_{-1.31}$)$\times10^{9}$ & $-$ & $-$ \\
%oz27
LBG\,J100111.03+022043.4 & 2.467 & UV & (5.69$^{+0.77}_{-0.68}$)$\times10^{10}$   & 45.0$^{+21.0}_{-17.0}$ & (1.65$^{+0.03}_{-0.43}$)$\times10^{10}$ & $-$ & $-$ \\

%ZCOS24
LBG\,J100100.91+021927.3 & 2.469 & UV & (5.07$^{+0.75}_{-0.65}$)$\times10^{10}$   & 32.3$^{+2.4}_{-7.9}$   & (3.46$^{+0.14}_{-1.19}$)$\times10^{9}$ & $-$ & $-$ \\

%lz30
LBG\,J095956.93+022118.5 & 2.470 & UV & (2.48$^{+0.65}_{-0.51}$)$\times10^{10}$   & 15.7$^{+0.5}_{-7.8}$   & (1.77$^{+0.01}_{-0.62}$)$\times10^{9}$ & $-$ & $-$ \\
%COMP971
LBG\,J100059.45+021957.4  & 2.470 & \ha/\lya\ & (3.75$^{+0.72}_{-0.60}$)$\times10^{10}$   & 16.9$^{+1.7}_{-4.2}$   & (2.66$^{+0.02}_{-0.64}$)$\times10^{10}$ & $-$ & Opt/X-ray \\
%ZCOS25
LBG\,J100100.91+021728.1 & 2.470 & \lya\ & (7.18$^{+0.84}_{-0.75}$)$\times10^{10}$   & 30.7$^{+0.1}_{-4.7}$   & (2.84$^{+0.11}_{-0.51}$)$\times10^{9}$ & $-$ & $-$ \\
%DSFG850.44
DSFG\,J100018.17+022250.4 & 2.470 & \ha\ & (1.01$^{+0.09}_{-0.08}$)$\times$10$^{11}$ & 365$^{+223}_{-138}$    & (1.91$^{+0.15}_{-0.41}$)$\times10^{10}$ & Merg & $-$ \\

%ZCOS02
LBG\,J100014.24+022516.7 & 2.471 & UV & (3.24$^{+0.65}_{-0.54}$)$\times10^{10}$   & 46.7$^{+1.8}_{-5.3}$   & (1.13$^{+0.05}_{-0.20}$)$\times10^{10}$ & Sph/Int & $-$ \\
%oz28
LBG\,J100115.18+022349.7 & 2.471 & \lya/UV & (3.48$^{+0.70}_{-0.59}$)$\times10^{10}$   & 17.8$^{+1.0}_{-6.7}$   & (5.54$^{+1.05}_{-7.27}$)$\times10^{10}$ & $-$ & Radio \\

%DSFG450.09
DSFG\,J100016.57+022638.4 & 2.472 & \ha & (8.82$^{+5.45}_{-3.37}$)$\times$10$^{9}$  & 714$^{+210}_{-162}$    & (1.09$^{+0.02}_{-0.34}$)$\times10^{11}$ & Disk & Radio \\
%DSFG450.28
DSFG\,J100056.83+022013.3 & 2.472 & \ha\ & (3.71$^{+0.69}_{-0.58}$)$\times$10$^{10}$ & 196$^{+88}_{-61}$      & (8.13$^{+0.76}_{-0.74}$)$\times10^{10}$ & $-$ & $-$ \\
%%lz37
%LBG\,J095940.95+022522.1 & 2.472 & UV & (4.28$^{+0.74}_{-0.63}$)$\times10^{10}$   & 21.5$^{+15.5}_{-2.4}$  & (4.95$^{+0.83}_{-2.21}$)$\times10^{9}$ & $-$ & $-$ \\
%ZCOS05
LBG\,J100018.04+021808.6 & 2.472 & UV & (4.58$^{+0.72}_{-0.62}$)$\times10^{10}$   & 71.7$^{+8.5}_{-17.1}$  & (7.44$^{+0.67}_{-1.61}$)$\times10^{9}$ & Sph/Int & $-$ \\
%ZCOS08
LBG\,J100020.50+022421.5 & 2.472 & UV & (4.37$^{+0.33}_{-0.31}$)$\times10^{10}$   & 33.1$^{+14.3}_{-7.5}$  & (1.94$^{+0.21}_{-0.56}$)$\times10^{10}$ & Sph & $-$ \\

%oz26
LBG\,J100109.29+022221.5 & 2.473 & \lya/UV & (3.73$^{+0.65}_{-0.56}$)$\times10^{10}$   & 52.6$^{+10.1}_{-27.8}$ & (1.04$^{+0.02}_{-0.26}$)$\times10^{10}$ & $-$ & $-$ \\

%ZCOS03
LBG\,J100015.38+022448.3 & 2.474 & \lya/[C{\sc iv}] & (4.93$^{+0.70}_{-0.62}$)$\times10^{10}$   & 11.8$^{+5.0}_{-1.9}$   & (1.32$^{+0.01}_{-0.37}$)$\times10^{10}$ & Sph & UV \\
%ZCOS00
LBG\,J100033.20+022225.0 & 2.474 & \lya/UV & (2.48$^{+0.60}_{-0.48}$)$\times10^{10}$   & 63.2$^{+15.6}_{-8.9}$  & (1.52$^{+0.04}_{-0.34}$)$\times10^{10}$ & Sph & $-$ \\

%lz33
LBG\,J100008.88+023044.1 & 2.475 & UV & (4.45$^{+0.72}_{-0.62}$)$\times10^{10}$   & 16.2$^{+5.0}_{-1.8}$   & (1.93$^{+0.19}_{-0.35}$)$\times10^{10}$ & $-$ & $-$ \\
%lz34
LBG\,J100012.37+023707.6 & 2.475 & UV & (3.05$^{+0.67}_{-0.55}$)$\times10^{10}$   & 131$^{+16}_{-32}$      & (2.16$^{+0.01}_{-0.32}$)$\times10^{10}$ & $-$ & $-$ \\
%ZCOS04
LBG\,J100015.87+021939.5 & 2.475 & UV & (3.31$^{+0.70}_{-0.58}$)$\times10^{10}$   & 79.4$^{+12.2}_{-21.2}$ & (3.15$^{+0.19}_{-0.80}$)$\times10^{10}$ & Disk/Int & $-$ \\
%ZCOS14
LBG\,J100025.28+022643.3 & 2.475 & UV & (4.58$^{+0.69}_{-0.60}$)$\times10^{10}$   & 45.2$^{+1.8}_{-8.1}$   & (2.01$^{+0.10}_{-0.78}$)$\times10^{10}$ & Merg & $-$ \\
%oz29
LBG\,J100116.15+021854.2 & 2.475 & UV & (3.86$^{+0.65}_{-0.56}$)$\times10^{10}$   & 23.3$^{+1.3}_{-12.5}$  & (4.99$^{+1.75}_{-1.23}$)$\times10^{9}$ & $-$ & $-$ \\

%%lz31
LBG\,J100002.03+023012.9 & 2.477 & \lya/UV & (5.46$^{+0.75}_{-0.66}$)$\times10^{10}$   & 23.3$^{+1.9}_{-14.0}$  & (4.38$^{+0.53}_{-0.70}$)$\times10^{9}$ & $-$ & $-$ \\
%ZCOS21
LBG\,J100054.07+022104.4 & 2.478 & \lya/[C{\sc iv}] & (4.37$^{+0.74}_{-0.63}$)$\times10^{10}$   & 22.9$^{+2.8}_{-4.5}$   & (2.00$^{+0.07}_{-0.49}$)$\times10^{10}$ & $-$ & UV/X-ray \\
%SPIRE457
DSFG\,J100026.73+022411.3 & 2.478 & \ha\ & (7.55$^{+6.21}_{-3.41}$)$\times$10$^{9}$  & 284$^{+179}_{-110}$    & (1.64$^{+0.01}_{-0.45}$)$\times10^{10}$ & Disk & Opt \\
%ZCOS12
LBG\,J100024.21+022741.3 & 2.479 & UV & (5.47$^{+0.77}_{-0.67}$)$\times10^{10}$   & 18.0$^{+0.2}_{-2.8}$   & (1.04$^{+0.04}_{-0.11}$)$\times10^{10}$ & Sph & $-$ \\
%COLDz
DSFG\,J100018.21+023456.7 & 2.479 & CO(1-0) & $<$3.0$\times$10$^{10}$                & 400$^{+165}_{-117}$    & (2.13$^{+0.02}_{-0.56}$)$\times10^{11}$ & $-$ & $-$ \\
%ZCOS09
QSO\,J100021.96+022356.7 & 2.480  &  \lya\ & (2.62$^{+0.14}_{-0.13}$)$\times10^{11}$ & ...                   & (6.59$^{+0.21}_{-0.86}$)$\times10^{10}$ & QSO/Sph & UV/X-ray \\
%DSFG850.20
DSFG\,J100027.14+023140.8 & 2.483 & \ha & (6.04$^{+0.71}_{-0.59}$)$\times$10$^{10}$ & 1090$^{+180}_{-160}$   & (4.10$^{+0.33}_{-0.92}$)$\times10^{10}$ & Sph & Opt \\

%%lz32
LBG\,J100004.33+022654.1 & 2.480 & UV & (6.22$^{+0.76}_{-0.68}$)$\times10^{10}$   & 37.9$^{+3.6}_{-8.3}$   & (4.03$^{+0.24}_{-0.66}$)$\times10^{9}$ & $-$ & $-$ \\
%lz36
LBG\,J100033.91+022713.2 & 2.481 & UV & (2.17$^{+0.64}_{-0.50}$)$\times10^{10}$   & 27.3$^{+0.5}_{-3.6}$   & (2.57$^{+0.14}_{-0.25}$)$\times10^{9}$ & $-$ & $-$ \\
%lz35
LBG\,J100023.31+023537.5 & 2.487 & UV & (3.69$^{+0.86}_{-0.70}$)$\times10^{10}$   & 27.9$^{+0.3}_{-11.8}$  & (3.96$^{+0.78}_{-0.83}$)$\times10^{9}$ & $-$ & $-$ \\
\hline\hline
\end{tabular}
}
\end{center}
{\small Source names indicate the selection method: submillimeter
  selection by DSFG, optical by LBG, and one unequivocal quasar by
  QSO. L$_{\rm UV}$ is the rest-frame ultraviolet luminosity computed
  from an extrapolated apparent magnitude at rest-frame 1600\AA.
{\sc Morph Class} refers to the CANDELS visual classification scheme
applied to {\it HST} H-band imaging.  'Sph' refers to spheroids,
'Disk' refers to disk-like galaxies, `QSO' refers to an unresolved
point source, `Merg' refers to a galaxy merger, and 'Int' refers to
some signature of galaxy interactions taking place.  }
\label{tab:allmems}
\end{table*}

\clname\ sweeps out an effective area of 200\,arcmin$^2$ (distributed
over an area extended 25\arcmin$\times$25\arcmin) and 28\,Mpc comoving
along the line of sight, and a volume $\sim$15000\,Mpc$^3$ comoving
(or 400\,Mpc$^3$ proper).  The range of redshifts $\Delta z=0.0239$
translates to a total end-to-end line-of-sight velocity range of
$\Delta v=2080\,$km\,s$^{-1}$.  Figure~\ref{fig:map} maps the
structure.  Note that another over-dense structure
\citep{chiang14a,chiang15a,diener15a} at $z=2.44-2.45$ sits nearby (at
distances of $\sim$50--100\,Mpc).

This structure also hosts an over-density of luminous active galactic
nuclei (AGN), an important signature of accreting supermassive black
holes.  Four galaxies (9.5\%) are luminous X-ray sources in this
structure \citep[at $L_{\rm
    X}>$10$^{43.7}$\,erg\,s$^{-1}$;][]{civano12a}, a factor of
21 higher than the expected volume density of AGN of
similarly high luminosities \citep{silverman08a}. Even among the seven
DSFGs, four exhibit unequivocal AGN characteristics in either the
X-ray, radio, optical or ultraviolet (57\%): a fraction nearly twice
as high as expected from previous work on AGN in non-clustered DSFGs
\citep{alexander05a}.  The depth of the {\it Chandra}-COSMOS
observations is only sufficient to detect the most X-ray luminous AGN,
so we stack the undetected population to search for possible
low-luminosity AGN, but found no X-ray emission.

One of the seven DSFGs and one member LBG host radio-loud AGN
\citep{schinnerer07a}.  DSFG\,J100016.57+022638.4 has a radio
luminosity of $L_{\rm 178MHz}=1.9\times10^{27}$\,W\,Hz$^{-1}$, nearly
analogous to a local Fanaroff-Riley class II
AGN \citep{fanaroff74a}.  It has the potential
to become the brightest cluster galaxy (BCG) when the structure
matures to $z\sim0$, consistent with other high-$z$ 
 over-densities that host single
radio-loud quasars \citep{venemans07a}.

Beyond X-ray and radio signatures of AGN, we find that three DSFGs and
one LBG either have very broad H$\alpha$ lines or high [N{\sc
    ii}]/\ha\ ratios consistent with AGN.  In summary, this structure
hosts 6 luminous AGN, 9 galaxies with strong signatures of AGN, 7
submillimeter-luminous DSFGs, for a total of 12 exceptionally rare
galaxies.

\section{Results}\label{sec:results}

\subsection{Control Galaxies outside of \clname.}

To draw comparisons between \clname's member galaxies and
field galaxies, we define a set of spectroscopically-confirmed
galaxies that {\it lack} close physical associations.  The
redshift range of the control sample is restricted to $\pm200$\,Myr of
the observed structure ($2.30<z<2.66$) to prevent
confusion between redshift evolution and environmentally-driven
differences.  
Of the 1072 $z$COSMOS sources which satisfy $2.30<z<2.66$, we remove
sources within 20\,Mpc of \clname\ (a factor of two in comoving
distance beyond the boundary of the structure itself) and any sources
with more than two neighbors within 5\,Mpc.  The resulting sample
consists of 401 galaxies; the vast majority of these are selected via
optical-color and appear to only differ with the \clname\ LBGs by
environment.

\subsection{Stellar Masses \&\ Star-Formation Rates.}

We use the MAGPHYS spectral energy distribution (SED) code
\citep{da-cunha08a} to constrain the UV-through-far-infrared SED
empirically using energy balance techniques.  We use a stellar
synthesis template library as input, and attenuation is 
determined from a mix of hot/cool/PAH dust grains.  Our interest
in using MAGPHYS is threefold: to measure star-formation rates for
galaxies not directly detected in the far-infrared, to estimate
stellar masses for all galaxies (comparing stellar mass estimates of
DSFGs with other techniques), and to compare physical properties of
\clname\ members with field galaxies.  
One galaxy,
QSO\,J100021.96$+$022356.7, lacks a SFR estimate due to lack of
far-infrared detection or disambiguation of quasar-dominated optical
SED from stellar emission.

Star formation rates for DSFGs are measured by fitting simple modified
blackbodies \citep[plus mid-infrared powerlaws, $\beta=1.8$,
  $\alpha=2.0$;][]{casey12a} to all available
far-infrared/submillimeter photometry ({\it Spitzer}, {\it Herschel}, \scubaii).
We integrate under the SED from 8--1000\um\ to get the total infrared
luminosity, $L_{\rm IR}$ and convert to a star-formation rate using
SFR/\sfr = 9.5$\times$10$^{-11}$\,L$_{\rm
  IR}$/\lsun\ \citep{kennicutt98b} adjusted for a Chabrier initial mass
function \citep{chabrier03a}.

LBGs in \clname\ are more massive in stars than control LBGs by a
factor of 1.5$\pm$0.3.  Given the uncertain star-formation histories
of DSFGs, we check that the MAGPHYS-generated stellar mass estimates
are not systematically biased by computing a rest-frame H-band
magnitude for each DSFG, removing any mid-infrared dusty power-law (on
average, $\sim$50\%\ contribution), and converting to stellar mass
using a range of appropriate mass-to-light ratios \citep[$L_{\rm
    H}/M_\star=7.9^{+0.6}_{-2.1}$\,\lsun/mag;][]{hainline11a}.  Two
DSFGs lack stellar mass estimates; DSFG\,J100016.57+022638.4 is
dominated by a mid-infrared powerlaw (with no constraint on the
underlying stellar emission) and DSFG\,J100027.14+023140.8 is blended
in the near-infrared with several nearby sources.  For the remainder,
we find consistency between the MAGPHYS-derived stellar masses and the
H-band derived stellar masses.

\subsection{Submillimeter Stacking.}

To investigate low-level submillimeter emission in the proto-cluster's
LBGs, we stack our 450\um\ and 850\um\ submillimeter maps at the
positions of known galaxies, both in \clname\ and in the control
sample.  Stacking analysis in the submillimeter requires the removal
of bright, significantly-detected point sources \citep{webb03a} so
that real DSFGs do not bias the measurement.  Thirty-one of 34 LBG
members fall within the sensitive area of the submillimeter maps (71
of 401 control sample LBGs).  Flux densities are measured by
inverse-variance weighting \citep{viero13a,coppin15a}.  At 850\um, we
measure a flux density of $S_{\rm 850}=0.25\pm0.16$\,mJy for 
\clname\ LBGs, and 0.11$\pm$0.13\,mJy for control LBGs.  In
other words LBGs in \clname\ are brighter at 850\um\ by a factor of
2.3$\pm$3.0 (although consistent with equal flux density, the
likelihood of greater 850\um\ in \clname\ is 76\%).  At 450\um, we
measure a flux density of $S_{\rm 450}=0.41\pm0.85$\,mJy for LBGs in
\clname, and 1.66$\pm$0.69\,mJy for control LBGs.

Although low signal-to-noise due to the small number of coadded
sources, these measurements together are suggestive that the mass of
cold dust and interstellar medium (ISM) is potentially higher for
galaxies in the dense structure, despite their comparable
star-formation rates.  Since galaxies' ISM masses scale directly to
their gas masses \citep[with roughly constant dust-to-gas
  ratio;][]{scoville14a}, we deduce that the molecular gas reservoirs
of the structure's galaxies are probably deeper, thus their potential
for heightened star-formation relative to similar galaxies living
outside of it.  Follow-up molecular gas measurements are needed to
confirm this intriguing lead.

\subsection{Rest-frame Optical Morphologies.}

\begin{figure*}
\centering
\includegraphics[width=2.0\columnwidth]{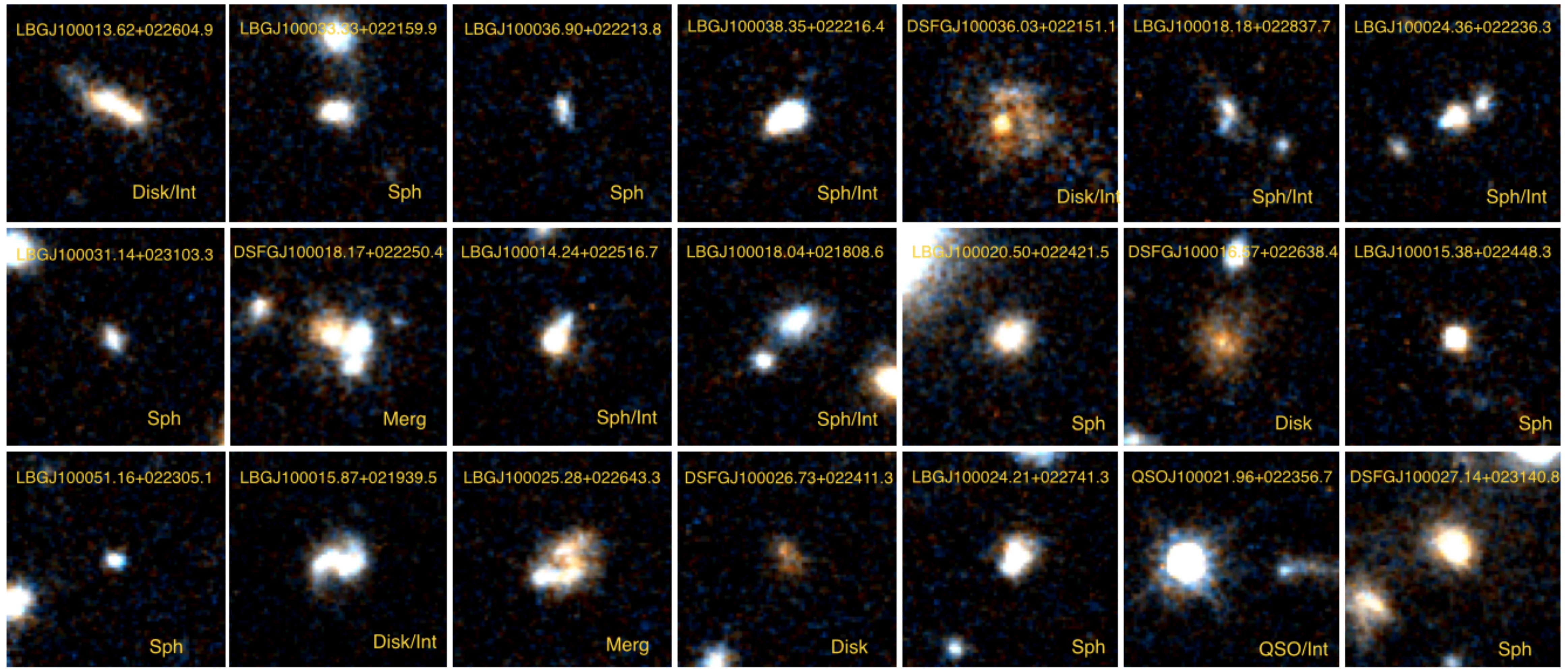}
\caption{$4\arcsec\times4\arcsec$ rest-frame optical cutouts for
  \clname\ members from {\it HST}-WFC3.  Galaxy names are indicated
  along with morphological and interaction indicators (see
  Table~\ref{tab:allmems}).}
\label{fig:cutouts}
\end{figure*}

With {\it Hubble Space Telescope (HST)} H-band imaging available for
21/42 \clname\ members (and 25/401 control galaxies), we investigate
morphology and interaction state of rest-frame optical emission using
the CANDELS visual classification scheme
\citep{kocevski12a,kartaltepe12a}.  The scheme classifies galaxies
into a morphology class (disk, spheroid, irregular, or unclassifiable)
and an interaction class (merger, interacting pair, or
non-interacting).  Full details of both morphology and interaction
class for member galaxies are given Table~\ref{tab:allmems} and
Figure~\ref{fig:cutouts}.  Before visual classification was carried
out, galaxy cutouts for cluster members and control were scrambled to
ensure unbiased results.  Although limited by small numbers, we find
that 10 of 21 proto-cluster member galaxies (48$\pm$10\%) appear to be irregular or
undergoing interaction while only 5 of
25 control galaxies exhibit interaction (20$\pm$8\%).  Even with
removal of the DSFGs, a high interaction fraction (7/16=44\%), is
found for the proto-cluster members.

\subsection{Estimating Halo Mass.}

To estimate the net dark matter halo mass of this structure, we use
abundance matching techniques from large-volume
simulations \citep*{behroozi13a}.  Due to \clname's filamentary
structure, we estimate the total halo mass by summing the estimated
halo masses for each galaxy in the structure using its stellar mass.
We estimate a lower limit of 
M$_{\rm halo}>$(8$\pm$3)$\times10^{13}$\,\msun\ at $z=2.47$.  Using a model for
mass-dependent exponential growth, we estimate the $z=0$ halo mass to
be (2$\pm$1)$\times10^{15}$\,\msun\ \citep{wechsler02a,chiang13a},
about twice as large as the Coma Supercluster (which has a mass of
$\sim$1$\times$10$^{15}$\msun).  Note that this dark matter halo mass
estimation method differs from others which 
assume linear bias and spherical collapse models
\citep{mo96a,peacock99a} which we suggest are not applicable to $z\simgt2$ filamentary structures.

Some works \citep{chapman09a,miller15a} suggest that significant
over-densities in DSFGs do not trace massive proto-clusters.  They
argue that such significant over-densities are due to ``merger
bias'' whereby the submillimeter-luminous phase is too short-lived and
rare to be a useful probe of the most massive halos at high-$z$.
Indeed, there are several massive proto-clusters at $z>2$ that contain
no DSFGs, and some structures of only moderate mass which appear to
contain a few DSFGs \citep{capak11a,hodge12a,walter12a}.  The
\citet{miller15a} work argues that DSFG over-densities are {\it poor}
tracers of the most massive over-densities at high-redshift because
Poisson noise dominates for low numbers of DSFGs.  While this is
likely the case for 1--3 DSFGs, our results (and our interpretation of
their Figure 3) imply that the opposite is actually true for
sufficiently large numbers of DSFGs per proto-cluster.  Instead we
suggest that spectroscopic incompleteness in both LBG and
submillimeter samples around high-$z$ proto-clusters has potentially
hindered the discovery of more starburst-enriched and AGN-enriched
proto-clusters.  If this is the case, aggressive spectroscopic
follow-up of DSFGs might substantially help the effort to identify
high-$z$ proto-cluster environments, where the spectroscopic
investment needed for LBG samples is prohibitive on large scales
(i.e. identification of $>$40 spectroscopically-identified LBGs, with
a resulting $\delta_{\rm LBG}\approx3-5$).  With the identification of
four 20--30 arcminute-scale over-densities containing $\ge$6 DSFGs
(HDF, \clname\ and SSA22) in only a few degrees of deep submillimeter
surveys, the potential to identify further massive cluster progenitors
via their member DSFGs and AGN is quite promising.

\section{Discussion \&\ Conclusions}\label{sec:conclusions}

The over-abundance of what are thought to be short-duration,
$\sim$100\,Myr, phenomena \citep[both DSFGs and luminous
  AGN;][]{bothwell13a,marconi04a} in an extended proto-cluster
structure is conspicuously rare.  
Even scaling density estimates proportional to the LBG
overenhancement, the DSFG/AGN presence is excessive.
The extra DSFG/AGN can be explained with only two possible
physical interpretations: either the DSFGs and AGN are short-lived and
triggered {\it simultaneously} via a process related to the over-dense
environment or the DSFGs and AGN must be much longer lived than
expected by existing observations of similar phenomena in the field.

The evidence we present here is suggestive of the former, that DSFGs
and AGN are short-lived, and in that case, their triggering must be
environmentally driven.  This is supported by the increased
interaction fraction seen in proto-cluster members' optical
morphologies, larger ISM masses in proto-cluster members (albeit a
marginal distinction requiring confirmation), and the lack of physical
motivation for long QSO lifetimes \citep{martini04a,hopkins09a} beyond
the increased gas-feeding argument often argued for DSFGs.
If correct, these observations provide the first concrete evidence
that environmental triggering can occur over extraordinarily large
volumes (15000\,Mpc$^3$ comoving) at $z>2$.

It should be noted that the structure's member LBGs do not
appear to have sufficiently different observational characteristics
than LBGs in the field.  Comparison against the control sample
indicates that proto-cluster LBGs have slightly higher stellar masses
(by a factor of 1.5$\pm$0.3) and similar star-formation rates.
Nevertheless, as a sheer consequence of their number, LBGs dominate
the calculation of the structure's net dark matter halo mass which is
estimated to be significant at $>$8$\times$10$^{13}$\,\msun.  This is predicted to
mature to a few $\times10^{15}$\,\msun at $z\sim0$.

Cosmological dark-matter simulations suggest that today's most massive
clusters occupied volumes several hundred times larger
\citep{onorbe14a} at $z\sim2.5$, having not yet virialized into the
compact structures we see today.  \clname\ affirms these predictions,
as its 15000\,Mpc$^3$ volume is predicted to collapse to a volume of
$\approx$\,50\,Mpc$^3$ at $z\sim0$, which is consistent with all
nearby $>5\times10^{14}$\,\msun\ clusters.  While this confirms the
notion of a genuine massive cluster in formation, this volume-scaling
also implies that {\it most} clusters will have similarly large sizes
at $z>2.5$, subtending areas half a degree across on the sky.
Observationally isolating massive clusters in formation then comes
down to accurate redshift identification to within $\Delta z=0.03$,
because other observational characteristics are not sufficiently
environmentally distinct at these epochs.

Identifying and correctly classifying \clname\ as a massive galaxy
cluster progenitor would not have been possible without the
concentrated presence of cosmologically rare phenomena like dusty
starbursts and luminous AGN.  
Future large and deep $\simgt$100\,deg$^2$ submillimeter
surveys could play a crucial part in statistically characterizing the
population of such large structures during their formation epoch, as
groups of DSFGs and luminous AGN can act as signposts to the largest
mass concentrations.  Equally important is complete spectroscopic
follow-up at $z>2$ over those wide-areas, like large
optical campaigns similar to HETDEX \citep{chiang14a} and
potential future large millimeter line searches
targeting CO or [C{\sc ii}] with a wide-bandwidth spectrometer. The
discovery of more high-$z$, starbursting over-densities will be
inevitable, but it will be the constraints on their volume density and
assembly timescale that will have significant repercussions on
cosmological hydrodynamic simulations and the formation mechanisms of
the Coma-like superclusters we see today.

\acknowledgements

We thank the anonymous referee for constructive comments which have
improved the manuscript.  Some of the data presented herein were
obtained at the W.M. Keck Observatory, which is operated as a
scientific partnership among the California Institute of Technology,
the University of California and the National Aeronautics and Space
Administration and made possible by financial support of the W.M. Keck
Foundation.  We also thank the Joint Astornomy Centre (run by STFC,
NSERC, NOSR) for operation of the JCMT and \scubaii\ instrument.
Mahalo nui loa to the kama'aina of Hawai'i for use of the cultural
site of Maunakea for astronomical observing.  Based in part on
NASA/ESA {\it HST} data, operated by STScI of AURA Inc, under NASA
contract NAS 5-26555. Other observations obtained from the European
Southern Observatory, Chile.  CMC acknowledges generous support of a
McCue Fellowship from the UC Irvine Center for Cosmology.  AC and CMC
acknowledge support from NSF AST-1313319, NSF CAREER 0645427 and NASA
Herschel Science Center.

%\bibliography{caitlin-bibdesk}

\begin{thebibliography}{50}
\expandafter\ifx\csname natexlab\endcsname\relax\def\natexlab#1{#1}\fi

\bibitem[{{Alexander} {et~al.}(2005){Alexander}, {Bauer}, {Chapman}, {Smail},
  {Blain}, {Brandt}, \& {Ivison}}]{alexander05a}
{Alexander}, D.~M., {Bauer}, F.~E., {Chapman}, S.~C., {Smail}, I., {Blain},
  A.~W., {Brandt}, W.~N., \& {Ivison}, R.~J. 2005, \apj, 632, 736

\bibitem[{{Behroozi} {et~al.}(2013){Behroozi}, {Wechsler}, \&
  {Conroy}}]{behroozi13a}
{Behroozi}, P.~S., {Wechsler}, R.~H., \& {Conroy}, C. 2013, \apj, 770, 57

\bibitem[{{Blain} {et~al.}(2004){Blain}, {Chapman}, {Smail}, \&
  {Ivison}}]{blain04a}
{Blain}, A.~W., {Chapman}, S.~C., {Smail}, I., \& {Ivison}, R. 2004, \apj, 611,
  725

\bibitem[{{Bothwell} {et~al.}(2013){Bothwell}, {Smail}, {Chapman}, {Genzel},
  {Ivison}, {Tacconi}, {Alaghband-Zadeh}, {Bertoldi}, {Blain}, {Casey}, {Cox},
  {Greve}, {Lutz}, {Neri}, {Omont}, \& {Swinbank}}]{bothwell13a}
{Bothwell}, M.~S., \etal 2013, \mnras, 429, 3047

\bibitem[{{Capak} {et~al.}(2011){Capak}, {Riechers}, {Scoville}, {Carilli},
  {Cox}, {Neri}, {Robertson}, {Salvato}, {Schinnerer}, {Yan}, {Wilson}, {Yun},
  {Civano}, {Elvis}, {Karim}, {Mobasher}, \& {Staguhn}}]{capak11a}
{Capak}, P.~L., \etal 2011, \nat, 470, 233

\bibitem[{{Casey}(2012)}]{casey12a}
{Casey}, C.~M. 2012, \mnras, 425, 3094

\bibitem[{{Casey} {et~al.}(2013){Casey}, {Chen}, {Cowie}, {Barger}, {Capak},
  {Ilbert}, {Koss}, {Lee}, {Le Floc'h}, {Sanders}, \& {Williams}}]{casey13a}
{Casey}, C.~M., \etal\ 2013, \mnras, 436, 1919

\bibitem[{{Casey} {et~al.}(2014){Casey}, {Narayanan}, \& {Cooray}}]{casey14a}
{Casey}, C.~M., {Narayanan}, D., \& {Cooray}, A. 2014, \physrep, 541, 45

\bibitem[{{Chabrier}(2003)}]{chabrier03a}
{Chabrier}, G. 2003, \pasp, 115, 763

\bibitem[{{Chapman} {et~al.}(2009){Chapman}, {Blain}, {Ibata}, {Ivison},
  {Smail}, \& {Morrison}}]{chapman09a}
{Chapman}, S.~C., {Blain}, A., {Ibata}, R., {Ivison}, R.~J., {Smail}, I., \&
  {Morrison}, G. 2009, \apj, 691, 560

\bibitem[{{Chiang} {et~al.}(2013){Chiang}, {Overzier}, \&
  {Gebhardt}}]{chiang13a}
{Chiang}, Y.-K., {Overzier}, R., \& {Gebhardt}, K. 2013, \apj, 779, 127

\bibitem[{{Chiang} {et~al.}(2014){Chiang}, {Overzier}, \&
  {Gebhardt}}]{chiang14a}
---. 2014, \apjl, 782, L3

\bibitem[{{Chiang} {et~al.}(2015){Chiang}, {Overzier}, {Gebhardt},
  {Finkelstein}, {Chiang}, {Hill}, {Blanc}, {Drory}, {Chonis}, {Zeimann},
  {Hagen}, {Schneider}, {Jogee}, {Ciardullo}, \& {Gronwall}}]{chiang15a}
{Chiang}, Y.-K., \etal\ 2015, ArXiv e-prints

\bibitem[{{Civano} {et~al.}(2012){Civano}, {Elvis}, {Brusa}, {Comastri},
  {Salvato}, {Zamorani}, {Aldcroft}, {Bongiorno}, {Capak}, {Cappelluti},
  {Cisternas}, {Fiore}, {Fruscione}, {Hao}, {Kartaltepe}, {Koekemoer}, {Gilli},
  {Impey}, {Lanzuisi}, {Lusso}, {Mainieri}, {Miyaji}, {Lilly}, {Masters},
  {Puccetti}, {Schawinski}, {Scoville}, {Silverman}, {Trump}, {Urry},
  {Vignali}, \& {Wright}}]{civano12a}
{Civano}, F., \etal\ 2012, \apjs, 201, 30

\bibitem[{{Cooper} {et~al.}(2008){Cooper}, {Tremonti}, {Newman}, \&
  {Zabludoff}}]{cooper08a}
{Cooper}, M.~C., {Tremonti}, C.~A., {Newman}, J.~A., \& {Zabludoff}, A.~I.
  2008, \mnras, 390, 245

\bibitem[{{Coppin} {et~al.}(2015){Coppin}, {Geach}, {Almaini}, {Arumugam},
  {Dunlop}, {Hartley}, {Ivison}, {Simpson}, {Smith}, {Swinbank}, {Blain},
  {Bourne}, {Bremer}, {Conselice}, {Harrison}, {Mortlock}, {Chapman}, {Davies},
  {Farrah}, {Gibb}, {Jenness}, {Karim}, {Knudsen}, {Ibar}, {Micha{\l}owski},
  {Peacock}, {Rigopoulou}, {Robson}, {Scott}, {Stevens}, \& {van der
  Werf}}]{coppin15a}
{Coppin}, K.~E.~K., \etal\ 2015, \mnras, 446, 1293

\bibitem[{{da Cunha} {et~al.}(2008){da Cunha}, {Charlot}, \&
  {Elbaz}}]{da-cunha08a}
{da Cunha}, E., {Charlot}, S., \& {Elbaz}, D. 2008, MNRAS, 388, 1595

\bibitem[{{Dannerbauer} {et~al.}(2014){Dannerbauer}, {Kurk}, {De Breuck},
  {Wylezalek}, {Santos}, {Koyama}, {Seymour}, {Tanaka}, {Hatch}, {Altieri},
  {Coia}, {Galametz}, {Kodama}, {Miley}, {R{\"o}ttgering}, {Sanchez-Portal},
  {Valtchanov}, {Venemans}, \& {Ziegler}}]{dannerbauer14a}
{Dannerbauer}, H., \etal\ 2014, \aap, 570, A55

\bibitem[{{Diener} {et~al.}(2013){Diener}, {Lilly}, {Knobel}, {Zamorani},
  {Lemson}, {Kampczyk}, {Scoville}, {Carollo}, {Contini}, {Kneib}, {Le Fevre},
  {Mainieri}, {Renzini}, {Scodeggio}, {Bardelli}, {Bolzonella}, {Bongiorno},
  {Caputi}, {Cucciati}, {de la Torre}, {de Ravel}, {Franzetti}, {Garilli},
  {Iovino}, {Kova{\v c}}, {Lamareille}, {Le Borgne}, {Le Brun}, {Maier},
  {Mignoli}, {Pello}, {Peng}, {Perez Montero}, {Presotto}, {Silverman},
  {Tanaka}, {Tasca}, {Tresse}, {Vergani}, {Zucca}, {Bordoloi}, {Cappi},
  {Cimatti}, {Coppa}, {Koekemoer}, {L{\'o}pez-Sanjuan}, {McCracken}, {Moresco},
  {Nair}, {Pozzetti}, \& {Welikala}}]{diener13a}
{Diener}, C., \etal\ 2013, \apj, 765, 109

\bibitem[{{Diener} {et~al.}(2015){Diener}, {Lilly}, {Ledoux},
    {Zamorani}, {Bolzonella}, {Murphy}, {Capak}, {Ilbert}, \&
    {McCracken}}]{diener15a} {Diener}, C., \etal\ 2015, \apj, 802, 31

\bibitem[{{Eadie} {et~al.}(1971){Eadie}, {Drijard}, \& {James}}]{eadie71a}
{Eadie}, W.~T., {Drijard}, D., \& {James}, F.~E. 1971, {Statistical methods in
  experimental physics}

\bibitem[{{Elbaz} {et~al.}(2007){Elbaz}, {Daddi}, {Le Borgne}, {Dickinson},
  {Alexander}, {Chary}, {Starck}, {Brandt}, {Kitzbichler}, {MacDonald},
  {Nonino}, {Popesso}, {Stern}, \& {Vanzella}}]{elbaz07a}
{Elbaz}, D., \etal\  2007, \aap, 468, 33

\bibitem[{{Fanaroff} \& {Riley}(1974)}]{fanaroff74a}
{Fanaroff}, B.~L. \& {Riley}, J.~M. 1974, \mnras, 167, 31P

\bibitem[{{Hainline} {et~al.}(2011){Hainline}, {Blain}, {Smail}, {Alexander},
  {Armus}, {Chapman}, \& {Ivison}}]{hainline11a}
{Hainline}, L.~J., {Blain}, A.~W., {Smail}, I., {Alexander}, D.~M., {Armus},
  L., {Chapman}, S.~C., \& {Ivison}, R.~J. 2011, \apj, 740, 96

\bibitem[{{Hodge} {et~al.}(2012){Hodge}, {Carilli}, {Walter}, {de Blok},
  {Riechers}, {Daddi}, \& {Lentati}}]{hodge12a}
{Hodge}, J.~A., {Carilli}, C.~L., {Walter}, F., {de Blok}, W.~J.~G.,
  {Riechers}, D., {Daddi}, E., \& {Lentati}, L. 2012, \apj, 760, 11

\bibitem[{{Hopkins} \& {Hernquist}(2009)}]{hopkins09a}
{Hopkins}, P.~F. \& {Hernquist}, L. 2009, \apj, 698, 1550

\bibitem[{{Huchra} \& {Geller}(1982)}]{huchra82a}
{Huchra}, J.~P. \& {Geller}, M.~J. 1982, \apj, 257, 423

\bibitem[{{Ilbert} {et~al.}(2013){Ilbert}, {McCracken}, {Le F{\`e}vre},
  {Capak}, {Dunlop}, {Karim}, {Renzini}, {Caputi}, {Boissier}, {Arnouts},
  {Aussel}, {Comparat}, {Guo}, {Hudelot}, {Kartaltepe}, {Kneib}, {Krogager},
  {Le Floc'h}, {Lilly}, {Mellier}, {Milvang-Jensen}, {Moutard}, {Onodera},
  {Richard}, {Salvato}, {Sanders}, {Scoville}, {Silverman}, {Taniguchi},
  {Tasca}, {Thomas}, {Toft}, {Tresse}, {Vergani}, {Wolk}, \&
  {Zirm}}]{ilbert13a}
{Ilbert}, O., \etal\ 2013, \aap, 556, A55

\bibitem[{{Kartaltepe} {et~al.}(2012){Kartaltepe}, {Dickinson}, {Alexander},
  {Bell}, {Dahlen}, {Elbaz}, {Faber}, {Lotz}, {McIntosh}, {Wiklind}, {Altieri},
  {Aussel}, {Bethermin}, {Bournaud}, {Charmandaris}, {Conselice}, {Cooray},
  {Dannerbauer}, {Dav{\'e}}, {Dunlop}, {Dekel}, {Ferguson}, {Grogin}, {Hwang},
  {Ivison}, {Kocevski}, {Koekemoer}, {Koo}, {Lai}, {Leiton}, {Lucas}, {Lutz},
  {Magdis}, {Magnelli}, {Morrison}, {Mozena}, {Mullaney}, {Newman}, {Pope},
  {Popesso}, {van der Wel}, {Weiner}, \& {Wuyts}}]{kartaltepe12a}
{Kartaltepe}, J.~S., \etal\ 2012, \apj, 757, 23

\bibitem[{{Kennicutt}(1998)}]{kennicutt98b}
{Kennicutt}, Jr., R.~C. 1998, \araa, 36, 189

\bibitem[{{Kocevski} {et~al.}(2012){Kocevski}, {Faber}, {Mozena}, {Koekemoer},
  {Nandra}, {Rangel}, {Laird}, {Brusa}, {Wuyts}, {Trump}, {Koo}, {Somerville},
  {Bell}, {Lotz}, {Alexander}, {Bournaud}, {Conselice}, {Dahlen}, {Dekel},
  {Donley}, {Dunlop}, {Finoguenov}, {Georgakakis}, {Giavalisco}, {Guo},
  {Grogin}, {Hathi}, {Juneau}, {Kartaltepe}, {Lucas}, {McGrath}, {McIntosh},
  {Mobasher}, {Robaina}, {Rosario}, {Straughn}, {van der Wel}, \&
  {Villforth}}]{kocevski12a}
{Kocevski}, D.~D., 2012, \apj, 744, 148

\bibitem[{{Lee} {et~al.}(2013){Lee}, {Sanders}, {Casey}, {Scoville}, {Hung},
  {Le Floc'h}, {Ilbert}, {Aussel}, {Capak}, {Kartaltepe}, {Roseboom},
  {Salvato}, {Aravena}, {Berta}, {Bock}, {Oliver}, {Riguccini}, \&
  {Symeonidis}}]{lee13a}
{Lee}, N., \etal 2013, \apj, 778, 131

\bibitem[{{Lentati} {et~al.}(2015){Lentati}, {Wagg}, {Carilli}, {Riechers},
  {Capak}, {Walter}, {Aravena}, {da Cunha}, {Hodge}, {Ivison}, {Smail},
  {Sharon}, {Daddi}, {Decarli}, {Dickinson}, {Sargent}, {Scoville}, \& {Smol{\v
  c}{\'c}}}]{lentati15a}
{Lentati}, L., \etal\ 2015, \apj, 800, 67

\bibitem[{{Lilly} {et~al.}(2009){Lilly}, {Le Brun}, {Maier}, {Mainieri},
  {Mignoli}, {Scodeggio}, {Zamorani}, {Carollo}, {Contini}, {Kneib}, {Le
  F{\`e}vre}, {Renzini}, {Bardelli}, {Bolzonella}, {Bongiorno}, {Caputi},
  {Coppa}, {Cucciati}, {de la Torre}, {de Ravel}, {Franzetti}, {Garilli},
  {Iovino}, {Kampczyk}, {Kovac}, {Knobel}, {Lamareille}, {Le Borgne}, {Pello},
  {Peng}, {P{\'e}rez-Montero}, {Ricciardelli}, {Silverman}, {Tanaka}, {Tasca},
  {Tresse}, {Vergani}, {Zucca}, {Ilbert}, {Salvato}, {Oesch}, {Abbas},
  {Bottini}, {Capak}, {Cappi}, {Cassata}, {Cimatti}, {Elvis}, {Fumana},
  {Guzzo}, {Hasinger}, {Koekemoer}, {Leauthaud}, {Maccagni}, {Marinoni},
  {McCracken}, {Memeo}, {Meneux}, {Porciani}, {Pozzetti}, {Sanders},
  {Scaramella}, {Scarlata}, {Scoville}, {Shopbell}, \& {Taniguchi}}]{lilly09a}
{Lilly}, S.~J., \etal\ 2009, \apjs, 184, 218

\bibitem[{{Marconi} {et~al.}(2004){Marconi}, {Risaliti}, {Gilli}, {Hunt},
  {Maiolino}, \& {Salvati}}]{marconi04a}
{Marconi}, A., {Risaliti}, G., {Gilli}, R., {Hunt}, L.~K., {Maiolino}, R., \&
  {Salvati}, M. 2004, \mnras, 351, 169

\bibitem[{{Martini}(2004)}]{martini04a}
{Martini}, P. 2004, Coevolution of Black Holes and Galaxies, 169

\bibitem[{{Miller} {et~al.}(2015){Miller}, {Hayward}, {Chapman}, \&
  {Behroozi}}]{miller15a}
{Miller}, T.~B., {Hayward}, C.~C., {Chapman}, S.~C., \& {Behroozi}, P.~S. 2015,
  ArXiv e-prints

\bibitem[{{Mo} \& {White}(1996)}]{mo96a}
{Mo}, H.~J. \& {White}, S.~D.~M. 1996, \mnras, 282, 347

\bibitem[{{Moster} {et~al.}(2013){Moster}, {Naab}, \& {White}}]{moster13a}
{Moster}, B., {Naab}, T., \& {White}, S.~D.~M. 2013, MNRAS, 428, 312

\bibitem[{{O{\~n}orbe} {et~al.}(2014){O{\~n}orbe}, {Garrison-Kimmel}, {Maller},
  {Bullock}, {Rocha}, \& {Hahn}}]{onorbe14a}
{O{\~n}orbe}, J., {Garrison-Kimmel}, S., {Maller}, A.~H., {Bullock}, J.~S.,
  {Rocha}, M., \& {Hahn}, O. 2014, \mnras, 437, 1894

\bibitem[{{Peacock}(1999)}]{peacock99a}
{Peacock}, J.~A. 1999, Royal Society of London Philosophical Transactions
  Series A, 357, 133

\bibitem[{{Schinnerer} {et~al.}(2007){Schinnerer}, {Smol{\v c}i{\'c}},
  {Carilli}, {Bondi}, {Ciliegi}, {Jahnke}, {Scoville}, {Aussel}, {Bertoldi},
  {Blain}, {Impey}, {Koekemoer}, {Le Fevre}, \& {Urry}}]{schinnerer07a}
{Schinnerer}, E., \etal\ 2007, \apjs, 172, 46

\bibitem[{{Scoville} {et~al.}(2014){Scoville}, {Aussel}, {Sheth}, {Scott},
  {Sanders}, {Ivison}, {Pope}, {Capak}, {Vanden Bout}, {Manohar}, {Kartaltepe},
  {Robertson}, \& {Lilly}}]{scoville14a}
{Scoville}, N., \etal\ 2014, \apj, 783, 84

\bibitem[{{Silverman} {et~al.}(2008){Silverman}, {Green}, {Barkhouse}, {Kim},
  {Kim}, {Wilkes}, {Cameron}, {Hasinger}, {Jannuzi}, {Smith}, {Smith}, \&
  {Tananbaum}}]{silverman08a}
{Silverman}, J.~D., \etal\ 2008, \apj, 679, 118

\bibitem[{{Steidel} {et~al.}(1998){Steidel}, {Adelberger}, {Dickinson},
  {Giavalisco}, {Pettini}, \& {Kellogg}}]{steidel98a}
{Steidel}, C.~C., {Adelberger}, K.~L., {Dickinson}, M., {Giavalisco}, M.,
  {Pettini}, M., \& {Kellogg}, M. 1998, \apj, 492, 428

\bibitem[{{Venemans} {et~al.}(2007){Venemans}, {R{\"o}ttgering}, {Miley}, {van
  Breugel}, {de Breuck}, {Kurk}, {Pentericci}, {Stanford}, {Overzier}, {Croft},
  \& {Ford}}]{venemans07a}
{Venemans}, B.~P., \etal\ 2007, \aap, 461, 823

\bibitem[{{Viero} {et~al.}(2013){Viero}, {Wang}, {Zemcov}, {Addison},
  {Amblard}, {Arumugam}, {Aussel}, {B{\'e}thermin}, {Bock}, {Boselli}, {Buat},
  {Burgarella}, {Casey}, {Clements}, {Conley}, {Conversi}, {Cooray}, {De
  Zotti}, {Dowell}, {Farrah}, {Franceschini}, {Glenn}, {Griffin},
  {Hatziminaoglou}, {Heinis}, {Ibar}, {Ivison}, {Lagache}, {Levenson},
  {Marchetti}, {Marsden}, {Nguyen}, {O'Halloran}, {Oliver}, {Omont}, {Page},
  {Papageorgiou}, {Pearson}, {P{\'e}rez-Fournon}, {Pohlen}, {Rigopoulou},
  {Roseboom}, {Rowan-Robinson}, {Schulz}, {Scott}, {Seymour}, {Shupe}, {Smith},
  {Symeonidis}, {Vaccari}, {Valtchanov}, {Vieira}, {Wardlow}, \&
  {Xu}}]{viero13a}
{Viero}, M.~P., \etal\ 2013, \apj, 772, 77

\bibitem[{{Walter} {et~al.}(2012){Walter}, {Decarli}, {Carilli}, {Bertoldi},
  {Cox}, {da Cunha}, {Daddi}, {Dickinson}, {Downes}, {Elbaz}, {Ellis}, {Hodge},
  {Neri}, {Riechers}, {Weiss}, {Bell}, {Dannerbauer}, {Krips}, {Krumholz},
  {Lentati}, {Maiolino}, {Menten}, {Rix}, {Robertson}, {Spinrad}, {Stark}, \&
  {Stern}}]{walter12a}
{Walter}, F., \etal\ 2012, \nat, 486, 233

\bibitem[{{Webb} {et~al.}(2003){Webb}, {Eales}, {Foucaud}, {Lilly},
  {McCracken}, {Adelberger}, {Steidel}, {Shapley}, {Clements}, {Dunne}, {Le
  F{\`e}vre}, {Brodwin}, \& {Gear}}]{webb03a}
{Webb}, T.~M., \etal\ 2003, \apj, 582, 6

\bibitem[{{Wechsler} {et~al.}(2002){Wechsler}, {Bullock}, {Primack},
  {Kravtsov}, \& {Dekel}}]{wechsler02a}
{Wechsler}, R.~H., {Bullock}, J.~S., {Primack}, J.~R., {Kravtsov}, A.~V., \&
  {Dekel}, A. 2002, \apj, 568, 52

\end{thebibliography}

\end{document}